\documentclass{llncs}
\usepackage[title]{appendix}
\usepackage[utf8]{inputenc}
\usepackage{fancyhdr}
\usepackage{indentfirst}
\usepackage{setspace}
\usepackage{graphicx,url}
\usepackage{colortbl}
\usepackage{caption}
\usepackage{subcaption}
\usepackage[table,xcdraw]{xcolor}
\usepackage{subfig} 
\usepackage[square,sort,comma,numbers]{natbib}
\usepackage{enumerate} 
\usepackage{float}
\usepackage{wrapfig}

\begin{document}
\title{Identifying Stereotypes in the\\Online Perception of Physical Attractiveness}

%
\titlerunning{Stereotypes in the Online Perception of Physical Attractiveness}  
%
\author{Camila Souza Ara{\'u}jo\inst{1} \and Wagner Meira Jr.\inst{1} \and Virgilio Almeida\inst{1}}

\authorrunning{Camila Ara{\'u}jo et al.} 
%
\tocauthor{Camila Souza Ara{\'u}jo, Wagner Meira Jr, and Virgilio Almeida}
\institute{Universidade Federal de Minas Gerais, Belo Horizonte MG, Brazil,\\
\email{\{camilaaraujo, meira, virgilio\}@dcc.ufmg.br},\\ 
}

\maketitle

\begin{abstract}
Stereotyping can be viewed as oversimplified ideas about social groups. They can be positive, neutral or negative. The main goal of this paper is to identify stereotypes for female physical attractiveness in images available in the Web. We look at the search engines as possible sources of stereotypes.  We  conducted experiments on Google and Bing by querying the search engines for beautiful and ugly women. We then collect images and extract information of faces. We propose a methodology and apply it to analyze photos gathered from search engines to understand how race and age manifest in the observed stereotypes and how they vary according to countries and regions. Our findings demonstrate the existence of stereotypes for female physical attractiveness, in particular negative stereotypes about black women and positive stereotypes about white women in terms of beauty. We also found negative stereotypes associated with older women in terms of physical attractiveness. Finally, we have identified patterns of stereotypes that are common to groups of countries.
\end{abstract}

\section{Introduction}

Prejudice, discrimination and stereotyping often go hand-in-hand in the real world.  While stereotyping can be viewed as oversimplified ideas about social groups, discrimination refers to actions that threat groups of people unfairly or put them at a disadvantage with other groups.  Stereotypes can be positive, neutral or negatives. For example, tiger moms are considered a positive stereotype that refers to Asian-American mothers  that keep focus on achievement and performance in the education of their children.  However, negative stereotypes based on gender, religion, ethnicity, sexual orientation and age can be harmful, for they may foster bias and discrimination.  As a consequence, they can lead to actions against groups of people \cite{Cash1989, kay2015unequal}. 

Some  appearance stereotypes associated with women in the physical world  follow them in the online world.  A recent study by Kay et al \cite{kay2015unequal} shows a systematic under representation of women in image search results for occupations. This kind of stereotype affects people’s ideas about professional gender ratios in the real world and may create conditions for bias and discrimination.

In the past, television, movies, and magazines  have  played a significant role in the creation and dissemination of stereotypes related to the physical appearance or physical attractiveness of women \cite{Downs1985}. The concepts of beauty, ugly, young and old have been used to  create categories of cultural and social stereotypes.  The idealized images of beautiful women have contributed to created negative consequences such as eating disorders, low self esteem and job discrimination. These stereotypes have  been a serious problem among teenage girls.

With the ongoing growth of Internet and social media, people are constantly exposed to steady flows of news, information and subjective opinions of others about cultural trends, political  facts, economic ideas, social issues, etc.  In addition to  information that come from different sources, people use Google to obtain  answers and information in order to form their own opinion on various social issues. Every day, Google processes over 3.5 billion  search queries.  Google decides which of the billions of web pages are  included in the search results, and it also decides how to rank the results.  Google  provides images as the result of  queries. Thus, in order to understand the existence of  global stereotypes, we need to start by looking at the search engines, as possible sources of stereotypes. In this paper we focus our analysis on the following research questions:

\begin{itemize}
\item Can we identify stereotypes for female physical attractiveness in the images available in the Web? 
\item How do race and age manifest in the observed stereotypes? 
\item How do stereotypes vary according to countries and regions?
\end{itemize}

In our analyses, we look for patterns of women's physical features that are considered aesthetically pleasing or beautiful in different cultures. We also look at the reverse, i.e., patterns are considered aesthetically ugly \cite{hogarth1753}. In order to answer the research questions, we  conduct a series of  experiments on the two most popular  search engines, Google and Bing. We start the experimentation  by querying the search engines for beauty and ugly women. We then collect  the top 50 image search results for up to 42 different countries. Once we have verified the images, we use Face++, which is an online API that detects faces in a given photo.  Face++ infers information about each face in the photo such as age, race and gender. Its accuracy is known to be over 90\% \cite{bakhshi2014faces}. The images collected from Google and Bing, classified by Face++, form the datasets used to conduct the stereotype analyses. Based on the data we collected, we have the following observations, which are explained throughout the paper.

\begin{itemize}
\item 
we have observed the existence of both  negative stereotypes for black women and positive stereotypes for white women in terms of beauty;
\item 
we have noticed that there are negative stereotypes about older women in terms of physical attractiveness;
\item 
we have identified patterns of stereotypes that are common to groups of countries. For example, US and several Hispanic countries share negative stereotypes about black women, positive stereotypes for white women and almost neutral about Asian women.
\end{itemize}
The first step in solving a problem is to recognize that it does exist. Our findings demonstrate the existence of stereotypes for female physical attractiveness. An important way to fight gender and age discrimination is to discourage stereotypes.



\section{Related Work}
In this section we present some related work on characterization studies of search engines, bias and discrimination in the media, as well as physical attractiveness.

\paragraph{Characterization of search engines}
Because of its scope and impact power, Google has become an object of study in the field of digital media and  key  to understand how the results of queries affect people who use search engines. Previous studies investigated the existence of bias in specific scenarios. \cite{noble2013google} shows how racial and gender identities may be misrepresented, when, in this context, there is commercial interest. The result of a query to Google typically prioritizes some kind of advertisement, which should - ideally - be related to the query. But search engines are often biased, so it is important to assess how the result ranking is built and how it affects the access to information \cite{introna2000shaping}. Some more recent results argue that discriminating a certain group is inappropriate, since search engines are 'information environments' that may affect the perception and behavior of people \cite{kay2015unequal}. One example of such discrimination is, when searching the names of people with black last names, the higher likelihood of getting ads suggesting that these people were arrested, or face a problem with justice, even when it did not happen \cite{sweeney2013discrimination}. In this case, the search algorithm supposedly discriminates a certain group of people while looking for profit from advertising. \cite{umoja2012missed} has questioned the commercial search engines because the way they represent women, especially black women, and other marginalized groups, regardless of cultural issues. This behavior masks and perpetuate unequal access to social, political and economic life of some groups. 

\paragraph{Bias and discrimination in the media} Media influences people's perceptions about ethnic issues \cite{davis1999racial}. In the USA, media tends to propagate stereotypes that benefit dominant groups. Black men, for example, are often stereotyped as violent. Even though much of the black population does not agree with the way they are represented and believe that this construction is harmful, unpleasant or distasteful. Uber drivers who have African American last names tend to get more negative reviews. Just as black tenants have less chances of getting a vacancy at rented apartments on Airbnb site \cite{aabid2016on}.  In the medical scenario, because of false judgments, black patients may receive inferior treatment compared to the treatment given to white people \cite{hoffman2016racial}. Many health-care professionals believe in biological differences with respect to black and white people, for example, black skin to be more resistant.

\paragraph{Beauty as a concept}
The reasons why beauty standards exist and how they are built are topics that are broadly discussed from the biological and evolutionary point of view. In the book ``The Analysis of Beauty''\cite{hogarth1753} published in 1753, the author describes theories of visual beauty and grace.  For the authors in \cite{berghe1986skin} the aesthetic preference of the human beings is a case of \textit{gene-culture co-evolution}. In other words, our standards of beauty are shaped, simultaneously, by a genetic and cultural evolution. Other studies \cite{grammer2003Darwinian, fink2006visible} argue that the beauty standards are part of human evolution and therefore reinforce characteristics related to health, among other features that may reflect the search for more 'qualified' partners for reproduction. Some works are concerned to understand how, despite cultural differences, the concept of beauty seems to be built in the same way worldwide. Diverse ethnic groups agree consistently over the beauty of faces \cite{cunningham1995their}, although they disagree regarding the attractiveness of female bodies. It is even possible to indicate which features are most desirable: childish face  features for women - big eyes, small nose, etc. In \cite{coetzee2014cross}, the authors conclude that: people tend to agree more with respect to faces that are more familiar and in some cultures the skin tone is more important in the classification of beautiful people, but, in other cases, it is the face shape. In Computer Science, using methods of machine learning, it is possible to predict, 0.6 of correlation, a face attractiveness score, showing that it is possible for a machine to learn what is beautiful from the point of view of a human \cite{eisenthal2006facial}.


\section{Data gathering and analysis}
\label{sec:analysis}

In this section we describe the methodology used for characterizing stereotypes. We use a database of photos and information extracted from these photos, in particular features of the people portrayed. The first step of the methodology involves the data collection process: what and how to collect the data. Then we extract information about the collected photos, using computer vision algorithms to identify race, age and gender of the people in each picture. The second part of the research refers to the use of  the collected information to identify stereotypes.

\subsection{Data Gathering}
\label{sec:methodData}

Data gathering was carried through two search engines APIs for images: Google and Bing. Once gathered, we extract features from the photos using Face++\footnote{http://www.faceplusplus.com/}. 

\begin{figure}[!h]
        \centering
        \includegraphics[scale=0.32]{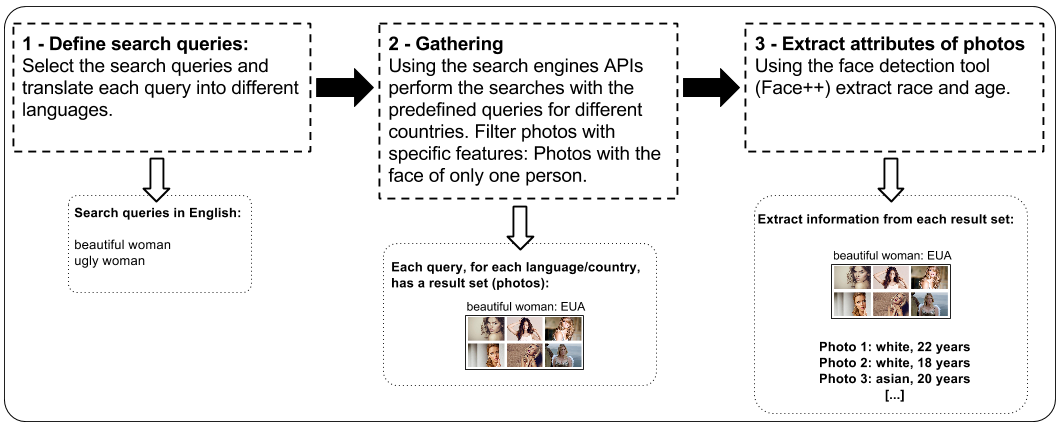}
   	\caption{Data Gathering Framework.}   
		\label{frame}
\end{figure}

The data gathering process is depicted in Figure~\ref{frame} and is summarized next:

\begin{small}
\begin{enumerate}
\item {\bf Define search queries}\\
For each context, in our case beauty, define the relevant search queries and translate the query to the target languages.
\item {\bf Gathering}\\Using the search engines APIs, perform the searches with the defined queries. Then, filter photos that contain the face of just one person.
\item {\bf Extract attributes of photos}\\Using face detection tools estimate race and age.
\end{enumerate}
\end{small}

Beauty is a property, or set of properties, that makes someone capable of producing a certain sort of pleasurable experience in any suitable perceiver \cite{TCDP1999}. For the beauty context we collected the top 50 photos of the results of the following queries (in different languages): beautiful woman and ugly woman.  It is known that what is defined as beautiful or ugly might change from person to person, then we chose these two antonyms adjectives that are commonly used to describe the quality of beauty of people.

Bing's API offers the option of 22 countries to perform the searches, we collected data for all these countries. For Google we collected data for the same 22 countries and added more countries with different characteristics, providing better coverage in terms of regions and internet usage. The searches were performed for the following countries and their official languages:

\begin{small}
\begin{description}
\item [Google:] Afghanistan, South Africa, Algeria, Angola, Saudi Arabia, Argentina, Australia, Austria, Brazil, Canada, Chile, Denmark, Egypt, Finland, France, Germany, Greece, Guatemala, India, Iraq, Ireland, Italy, Japan , Kenya, United Kingdom, South Korea, Malaysia, Mexico, Morocco, Nigeria, Paraguay, Peru, Portugal, Russia, Spain, United States, Sweden, Turkey, Ukraine, Uzbekistan, Venezuela and Zambia.
\item [Bing:] Saudi Arabia, Denmark, Austria, Germany, Greece, Australia, Canada, United Kingdom, United States, South Africa, Argentina, Spain, Mexico, Finland, Italy, Japan, South Korea, Brazil, Portugal, Russia, Turkey and Ukraine.
\end{description}
\end{small}

Now we present a brief characterization of the datasets collected for this work. As mentioned, we picked the top 50 photos for each query but we consider as valid only images for which Face++ was able to detect a single face (see appendix~\ref{app1}). The characterization and analysis will be performed for all query responses that contain at least 20 valid photos.

\begin{figure}[!h]
  \begin{subfigure}[t]{0.4\textwidth}
          \centering
          \includegraphics[scale=0.34]{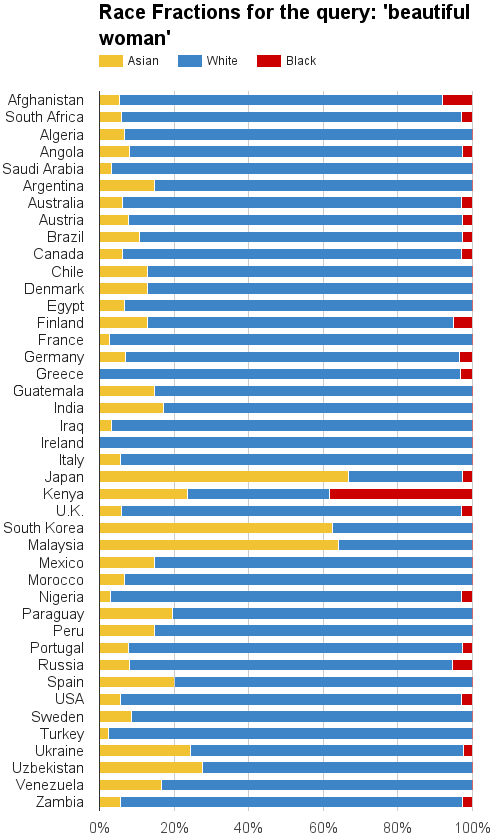}
          \label{propBeautyGoogle}
  \end{subfigure}
	\hspace{3em}
  \begin{subfigure}[t]{0.4\textwidth}
          \centering
          \includegraphics[scale=0.34]{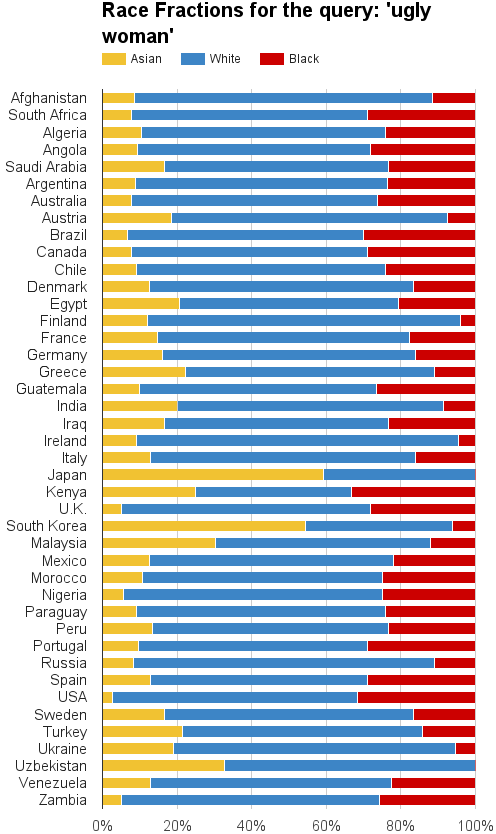}   
          \label{propUglyGoogle}
  \end{subfigure}
  \caption{Race Fractions for Google.} 
  \label{propGoogle}
\end{figure}

\begin{figure}[!h]
  \begin{subfigure}[t]{0.4\textwidth}
          \centering
          \includegraphics[scale=0.34]{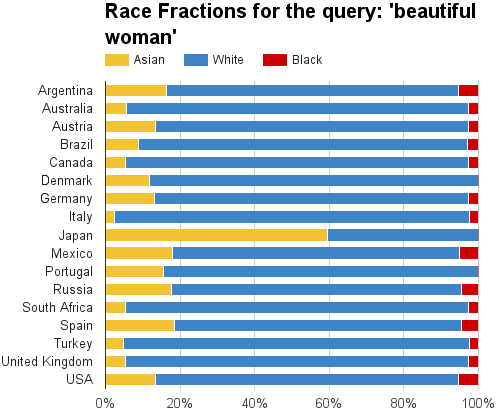}
          \label{propBeautyBing}
  \end{subfigure}
	\hspace{3em}
  \begin{subfigure}[t]{0.4\textwidth}
          \centering
          \includegraphics[scale=0.34]{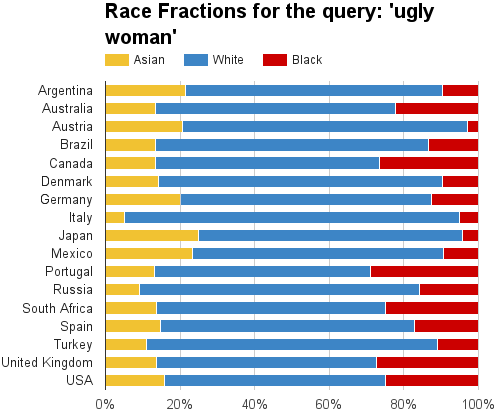}
          \label{propUglyBing}
  \end{subfigure}
  \caption{Race Fractions for Bing.} 
  \label{propBing}
\end{figure}

For the first step of the characterization our aim is to show the fraction of the races by country.  Figure~\ref{propGoogle} shows this fraction for the 42 countries for which we performed searches on Google and in Figure~\ref{propBing} for the 17 countries for Bing. Our first observation from the charts is that the fraction of black women in search 'ugly women' is clearly larger, in general, for the two search engines. We have also calculated the mean and standard deviation of each race for both queries and search engines. From the results in Table~\ref{tabMeanStDv} we observed the same for Asian women. 

\begin{table}[!h]
\centering
\begin{scriptsize}
\caption{Mean and Standard Deviation of Fractions}
\label{tabMeanStDv}
\begin{tabular}{cccccccccccc}
\hline
\multicolumn{12}{c}{\textbf{Google}}                                  \\ \hline
\multicolumn{6}{c|}{\textit{\textbf{beautiful woman}}}                                                                                                                     & \multicolumn{6}{c}{\textit{\textbf{ugly woman}}}                                                                                                      \\ \hline
\multicolumn{2}{c|}{\textit{Asian}}                     & \multicolumn{2}{c|}{\textit{Black}}                    & \multicolumn{2}{c|}{\textit{White}}                     & \multicolumn{2}{c|}{\textit{Asian}}                     & \multicolumn{2}{c|}{\textit{Black}}                    & \multicolumn{2}{c}{\textit{White}} \\ \hline
\multicolumn{1}{c|}{mean}  & \multicolumn{1}{c|}{stdv}  & \multicolumn{1}{c|}{mean}  & \multicolumn{1}{c|}{stdv} & \multicolumn{1}{c|}{mean}  & \multicolumn{1}{c|}{stdv}  & \multicolumn{1}{c|}{mean}  & \multicolumn{1}{c|}{stdv}  & \multicolumn{1}{c|}{mean}  & \multicolumn{1}{c|}{stdv} & \multicolumn{1}{c|}{mean}   & stdv \\ \hline
\multicolumn{1}{c|}{13.77} & \multicolumn{1}{c|}{15.65} & \multicolumn{1}{c|}{2.37}  & \multicolumn{1}{c|}{5.99} & \multicolumn{1}{c|}{83.86} & \multicolumn{1}{c|}{16.96} & \multicolumn{1}{c|}{15.36} & \multicolumn{1}{c|}{11.48} & \multicolumn{1}{c|}{19.20} & \multicolumn{1}{c|}{9.23} & \multicolumn{1}{c|}{65.44}  & 9.48 \\
                           &                            &                            &                           &                            &                            &                            &                            &                            &                           &                             &      \\ \hline
\multicolumn{12}{c}{\textbf{Bing}}                                                                                                                                                                                                                                                                                                 \\ \hline
\multicolumn{6}{c|}{\textit{\textbf{beautiful woman}}}                                                                                                                     & \multicolumn{6}{c}{\textit{\textbf{ugly woman}}}                                                                                                      \\ \hline
\multicolumn{2}{c|}{\textit{Asian}}                     & \multicolumn{2}{c|}{\textit{Black}}                    & \multicolumn{2}{c|}{\textit{White}}                     & \multicolumn{2}{c|}{\textit{Asian}}                     & \multicolumn{2}{c|}{\textit{Black}}                    & \multicolumn{2}{c}{\textit{White}} \\ \hline
\multicolumn{1}{c|}{mean}  & \multicolumn{1}{c|}{stdv}  & \multicolumn{1}{c|}{mean}  & \multicolumn{1}{c|}{stdv} & \multicolumn{1}{c|}{mean}  & \multicolumn{1}{c|}{stdv}  & \multicolumn{1}{c|}{mean}  & \multicolumn{1}{c|}{stdv}  & \multicolumn{1}{c|}{mean}  & \multicolumn{1}{c|}{stdv} & \multicolumn{1}{c|}{mean}   & stdv \\ \hline
\multicolumn{1}{c|}{12.96} & \multicolumn{1}{c|}{11.82} & \multicolumn{1}{c|}{03.09} & \multicolumn{1}{c|}{2.59} & \multicolumn{1}{c|}{83.94} & \multicolumn{1}{c|}{11.78} & \multicolumn{1}{c|}{15.35} & \multicolumn{1}{c|}{5.19}  & \multicolumn{1}{c|}{15.63} & \multicolumn{1}{c|}{8.54} & \multicolumn{1}{c|}{69.02}  & 5.19
\end{tabular}
\end{scriptsize}
\end{table}

\begin{figure}[!h]
  \begin{subfigure}[b]{0.55\textwidth}
          \centering
          \includegraphics[scale=0.23]{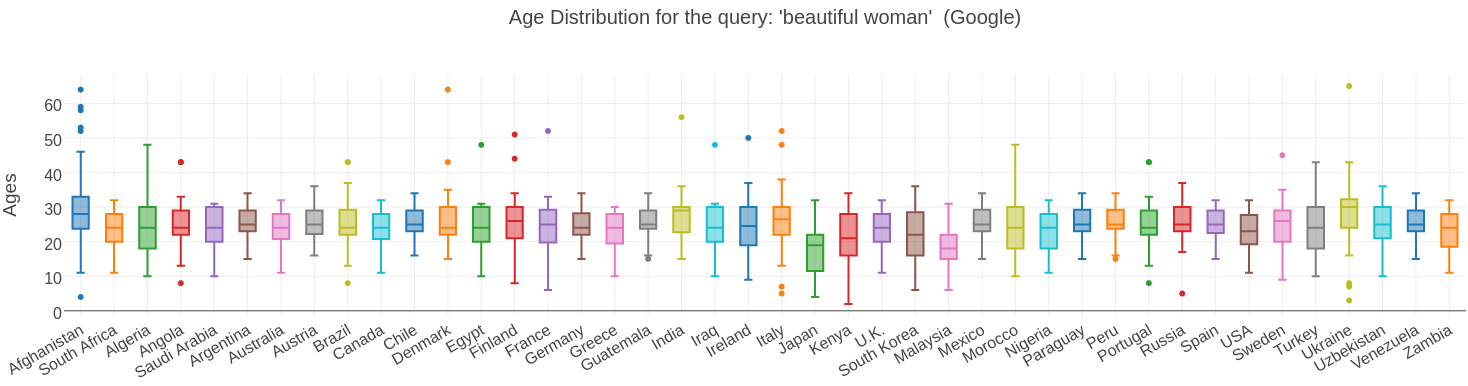}
          \label{distBeautyGoogle}
  \end{subfigure}
  \newline
  \begin{subfigure}[b]{0.55\textwidth}
          \centering
          \includegraphics[scale=0.23]{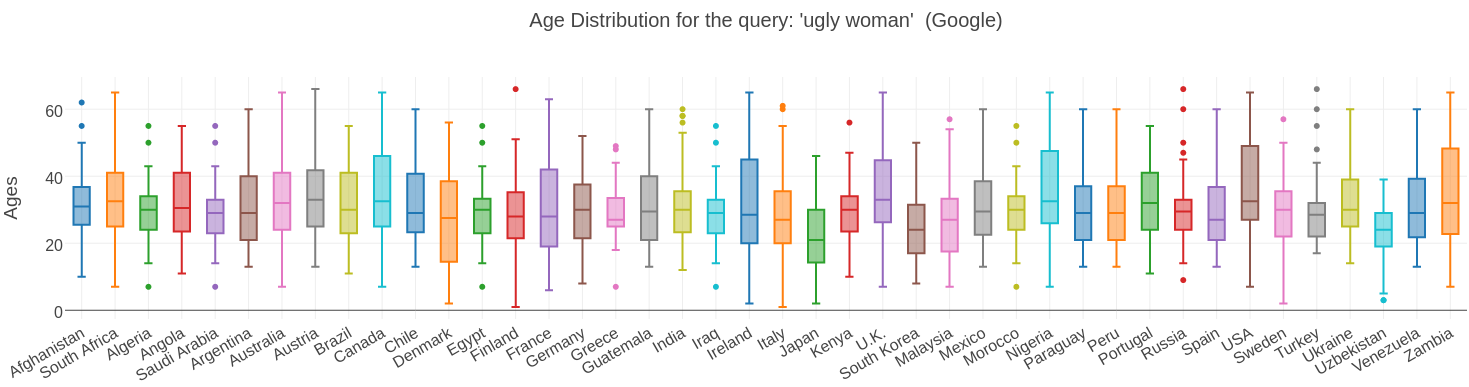}
          \label{distUglyGoogle}
  \end{subfigure}
  \caption{Age distribution for Google.} 
  \label{distGoogle}
\end{figure}

\begin{figure}[!h]
  \begin{subfigure}[b]{0.55\textwidth}
          \centering
          \includegraphics[scale=0.3]{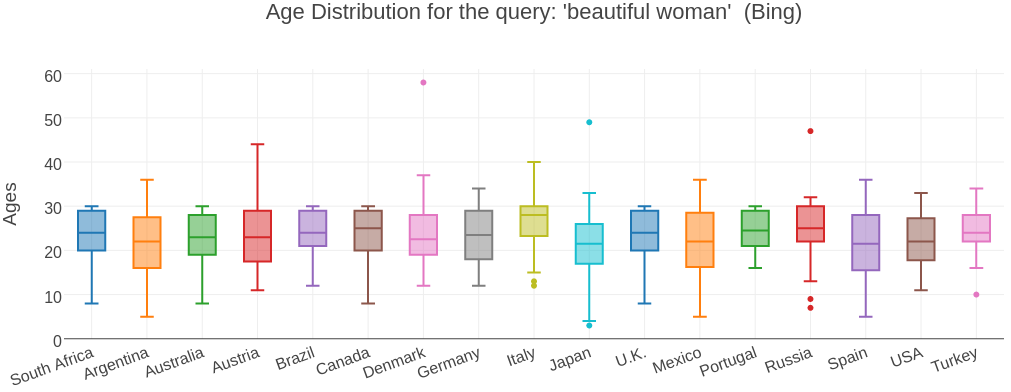}
          \label{distBeautyBing}
  \end{subfigure}
  \newline
  \begin{subfigure}[b]{0.55\textwidth}
          \centering
          \includegraphics[scale=0.3]{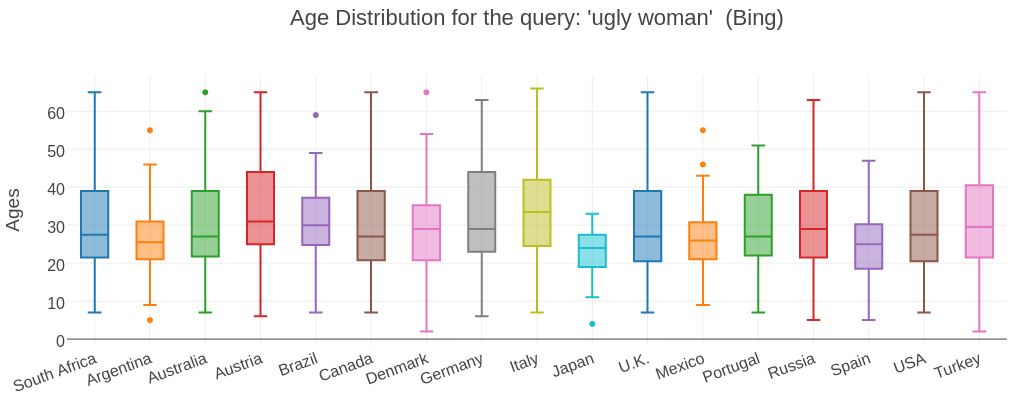}
          \label{distUglyBing}
  \end{subfigure}
  \caption{Age distribution for Bing.} 
  \label{distBing}
\end{figure}

The second step of the characterization shows the difference between the age distribution of women in photos by query and search engine through boxplots (Figures~\ref{distGoogle} and ~\ref{distBing}). In the x-axis we have the analyzed countries and the y-axis represents ages. Analyzing the median and upper quartile, we noticed that beautiful women tend to be younger than the ugly women.

\subsection{Data Analysis}
\label{sec:analysis}

Our main purpose is to identify whether there is a stereotype in the perception of physical attractiveness. For sake of our analysis, we distinguish two characteristics extracted from the pictures: race and age. As discussed, stereotype is a subjective concept and quantifying it through objective criteria is a challenge. In our case, we employed a {\em contrast}-based strategy. Considering race as a criteria, we check the difference between the fractions of each race for opposite queries, that is, beautiful woman and ugly woman. We consider that there is a negative stereotype of beauty in relation to a race, when the frequency of this particular race is larger when we search for ugly women compared to when we search for beautiful woman. Likewise, the stereotype is considered to be positive when the fraction is larger when we search for beautiful woman. Similarly, we say that there is a age stereotype when the age range of the women are younger in the searches for beautiful women. We characterize the occurrence of these stereotypes through seven questions:

\begin{small}
\begin{description}
\item [Q1:] Is the fraction of black women larger when we search for ugly women than when we search for beautiful women?
\item [Q2:] Is the fraction of Asian women larger when we search for ugly women than when we search for beautiful women?
\item [Q3:] Is the fraction of white women larger when we search for ugly women than when we search for beautiful women?
\item [Q4:] Is the fraction of black women smaller when we search for ugly women than when we search for beautiful women?
\item [Q5:] Is the fraction of Asian women smaller when we search for ugly women than when we search for beautiful women?
\item [Q6:] Is the fraction of white women smaller when we search for ugly women than when we search for beautiful women?
\item [Q7:] Are the women's ages when we search for beautiful women younger than the ages of the women when we search for ugly women?
\end{description}
\end{small}

Each of these questions is associated with a test hypothesis. For the questions \textbf{Q1}, \textbf{Q2} and \textbf{Q3}, the test hypothesis is:

\begin{small}
\begin{description}
\item [$H_{0}$(null hypothesis)]: The fraction of women of the specific race (i.e., black, white, Asian) is smaller when we search for ugly women, than when we search for beautiful women.
\item  [$H_{a}$(alternative hypothesis)]: The fraction of women of the specific race (i.e., black, white, Asian) is larger when we search for ugly women than when we search for beautiful women.
\end{description}
\end{small}

For the questions \textbf{Q4}, \textbf{Q5} and \textbf{Q6}:

\begin{small}
\begin{description}
\item [$H_{0}$]: The fraction of women of the specific race (black, white, Asian) is larger when we search for ugly women than when we search for beautiful women.
\item  [$H_{a}$]: The fraction of women of the specific race (black, white, Asian) is smaller when we search for ugly women than when we search for beautiful women.
\end{description}
\end{small}

For the question \textbf{Q7}:

\begin{small}
\begin{description}
\item [$H_{0}$]: The age range of the beautiful women is older than the age range of the ugly women.
\item  [$H_{a}$]: The age range of the beautiful women is younger than the age range of the ugly women.
\end{description}
\end{small}

We assume that there is a negative stereotype when the fraction of a given race is significantly larger when we search for ugly woman than when we search for beautiful woman and there is a positive stereotype when the fraction associated with a search for ugly woman is significantly smaller. We then calculate the difference between these two fractions for each race and each country and verify the significance of each difference through the {\bf two-proportion z-test}, with a significance level of $0.05$. This test determines whether the difference between the fractions is significant, as follows.

\subsubsection{Racial Stereotype}

For the first three questions, ({\bf Q1}, {\bf Q2} and {\bf Q3}), with confidence of 95\% we reject the null hypothesis when the z-score is smaller than $-0.8289$ and we accept the alternative hypothesis, which is the hypothesis in study. For example, considering Afghanistan, the z-score calculated for the hypothesis associated with question {\bf Q1} was $-0.48$, $-0.53 $ for {\bf Q2} and $0.74$ for {\bf Q3}. Since none of these values is smaller than $-0.8289$ we can not reject the null hypothesis and we can not answer positively to any of the 3 questions. On the other hand, for Italy, the z-score associated with question {\bf Q1} was $-2.51$ and $-1.05$ for {\bf Q2}, then we can answer positively to both questions and consider that there is a negative stereotype associated with blacks and Asians.

For questions ({\bf Q4}, {\bf Q5} and {\bf Q6}), under the same conditions, we reject the null hypothesis when the z-score is greater than $0.8289$. Detailed results of the tests and z-scores for each country and each search engine are in the appendix~\ref{app2}.

\subsubsection{Age Stereotype}

For characterizing the age stereotype, we verify our hypothesis through the unpaired Wilcoxon test~\cite{wilcoxon1945individual}. The null hypothesis is rejected when p-value is less than 0.05. and with 95\% of confidence we can answer positively to question {\bf Q7} (see appendix~\ref{app3} for detailed results). Once again, considering Afghanistan, the p-value found was $0.1819$ then we can not reject the null hypothesis. For South Africa the p-value was $0.0001$ and we accept the alternative hypothesis that demonstrates the existence of a stereotype that gives priority to younger women.

Table \ref{summary} summarizes the test results with the fraction of countries that we answer positively to each of the 7 questions (reject the null hypothesis). For instance, column 'Google' and line 'Q1' indicates that for $85.71\% $ of countries we rejected the null hypothesis and we answered positively to the question {\bf Q1}. That is, for almost $86\%$ of the countries the fraction of black women is larger when we search for ugly women than when we search for beautiful women. We can see that the results of the two search engines agree. There is a beauty stereotype in the perception of physical attractiveness, that is, we can say that significantly the fraction of black and Asian women is greater when we search for ugly women compared to the fraction of those races when we search for beautiful women (negative stereotype). The opposite occurs for white women (positive stereotype).

\begin{table}[!h]
\caption{Summary of results for questions {\bf Q1}, {\bf Q2}, {\bf Q3}, {\bf Q3}, {\bf Q4}, {\bf Q5}, {\bf Q6} e {\bf Q7}}
\label{summary}

\centering
\begin{tabular}{l|c|c|}
\cline{2-3}
                    & \multicolumn{2}{c|}{\textbf{Results}}                                   \\ \cline{2-3} 
                    & \multicolumn{1}{c|}{\textit{\textbf{Google}}} & \textit{\textbf{Bing}} \\ \cline{2-3} 
\hline   
\multicolumn{1}{|l|}{ \textit{Q1 (Black)}} & \multicolumn{1}{c|}{\textbf{85.71\%}}         & \textbf{76.47\%}       \\ \hline
\multicolumn{1}{|l|}{ \textit{Q2 (Asian)}} & \multicolumn{1}{c|}{\textbf{26.19\%}}         & \textbf{29.41\%}       \\ \hline
\multicolumn{1}{|l|}{ \textit{Q3 (White)}} & \multicolumn{1}{c|}{4.76\%}                   & 5.88\%                 \\ \hline
\multicolumn{1}{|l|}{ \textit{Q4 (Black)}} & \multicolumn{1}{c|}{2.38\%}                   & 0.00\%                 \\ \hline
\multicolumn{1}{|l|}{ \textit{Q5 (Asian)}} & \multicolumn{1}{c|}{4.76\%}                   & 11.76\%                \\ \hline
\multicolumn{1}{|l|}{ \textit{Q6 (White)}} & \multicolumn{1}{c|}{\textbf{78.57\%}}         & \textbf{82.35\%}       \\ \hline
\multicolumn{1}{|l|}{ \textit{Q7 (Age)}}   & \multicolumn{1}{c|}{\textbf{69.05\%}}         & \textbf{82.35\%} \\ \hline

\end{tabular}

\end{table}

\subsection{Clustering Stereotypes}
\label{sec:cluster}

After identifying the existence of stereotypes in the perception of physical attractiveness, we want to discover whether there is a cohesion among these beauty stereotypes across countries. For this we will use a clustering algorithm to identify the countries that have the same racial stereotype of beauty. The results for each country and search engine is represented by a 3D point where the dimensions are Asian, black and white z-scores.

There are several strategies for clustering, however a hierarchical clustering strategy was used in this paper because it outputs a hierarchy that can be very useful for our analysis. We used the Ward's minimum variance method \footnote{R library: https://stat.ethz.ch/R-manual/R-devel/library/stats/html/hclust.html} which is briefly described next. Using a set of dissimilarities for the objects being clustered, initially, each object is assigned to its own cluster and then the algorithm proceeds interactively. At each stage it joins the two most similar clusters, continuing until there is just a single cluster. The method aims at finding compact and spherical clusters\cite{murtagh2014ward}. Another advantage of employing a hierarchical clustering strategy is that it is not necessary to set in advance parameters such as the number of clusters of minimal similarity thresholds, allowing us to investigate various clusters configurations easily.

The clusters we are looking for should be cohesive and also semantically meaningful. Cohesion is achieved by the Ward's minimum variance method, but the semantic of the clusters should take into account cultural, political and historical aspects. In our case, the variance is taken in its classical definition, that is, it measures how far the entities, each one represented by a numeric triple (Q1, Q2 and Q6), that compose a cluster are spread out from their mean. For the results presented here we traversed the dendrogram starting from the smallest variance to the maximum variance, which is the root of the dendrogram. For each group of entities, we verify what they do have in common so that we may understand why they behaved similarly or not. As we show next, we are able to identify relevant and significant stereotypes across several entities (e.g., countries).

\begin{figure}[!h]
  \begin{subfigure}[b]{0.4\textwidth}
          \centering
        \includegraphics[scale=0.5]{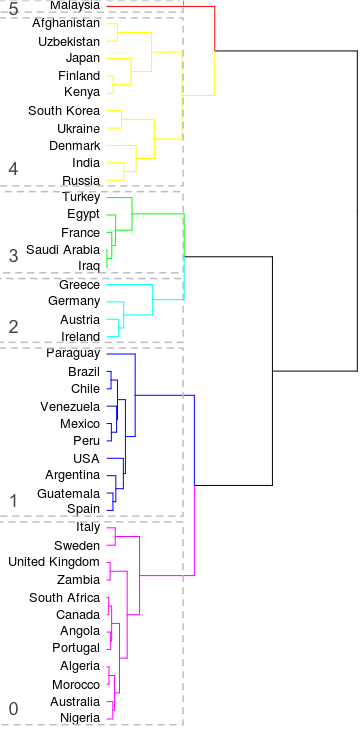}
   	\caption{Dendrogram with the cutoff of 6 clusters for Google.}  
		\label{clusteringGoogle}
  \end{subfigure}
  \hfill
  \begin{subfigure}[b]{0.4\textwidth}
          \centering
        \includegraphics[scale=0.45]{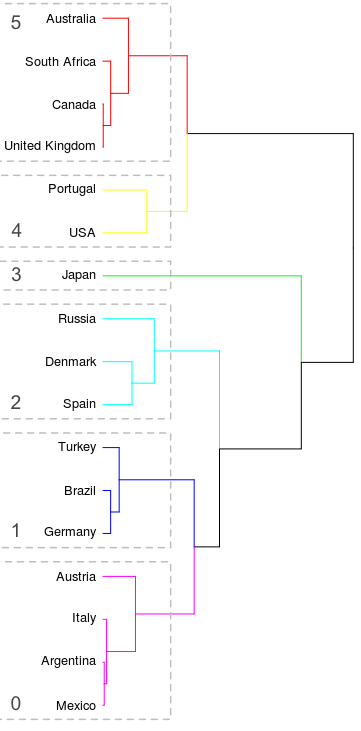}
   	\caption{Dendrogram with the cutoff of 6 clusters for Bing.}   
		\label{clusteringBing}
  \end{subfigure}
  \caption{Clusters} 
  \label{clusters}
\end{figure}

Figure~\ref{clusters} presents the dendrograms for both search engines, we use a cutoff of 6 clusters to illustrate the process of clustering from the dendrogram structure. The centroids of the clusters are shown in Table~\ref{tabCentroids}. It is important to emphasize that when analyzing the centroids of each cluster the dimensions represent the per race average z-score. In our previous analysis we have shown that for black and Asian women, a more negative score represents a stronger negative stereotype regarding the two races. For white women, a more positive score represents a stronger positive stereotype.

\begin{table}[!h]
\centering
\caption{Clusters centroids}
\label{tabCentroids}
\begin{scriptsize}
\begin{tabular}{lcccccc}
\cline{2-7}
                                         & \multicolumn{6}{c}{\textbf{Google}}                                                                                                                                                                             \\ \cline{2-7} 
\multicolumn{1}{l|}{}                    & \multicolumn{2}{c|}{\textit{\textbf{Black}}}                        & \multicolumn{2}{c|}{\textit{\textbf{Asian}}}                        & \multicolumn{2}{c|}{\textit{\textbf{White}}}                        \\ \cline{2-7} 
\multicolumn{1}{l|}{}                    & \textit{Mean}                 & \multicolumn{1}{c|}{\textit{STDEV}} & \textit{Mean}                 & \multicolumn{1}{c|}{\textit{STDEV}} & \textit{Mean}                 & \multicolumn{1}{c|}{\textit{STDEV}} \\ \hline
\multicolumn{1}{|l|}{\textit{Cluster 0}} & -2.85                         & \multicolumn{1}{c|}{0.16}           & -0.41                         & \multicolumn{1}{c|}{0.38}           & 2.65                          & \multicolumn{1}{c|}{0.12}           \\
\multicolumn{1}{|l|}{\textit{Cluster 1}} & -3.28                         & \multicolumn{1}{c|}{0.21}           & 0.60                          & \multicolumn{1}{c|}{0.31}           & 2.02                          & \multicolumn{1}{c|}{0.34}           \\
\multicolumn{1}{|l|}{\textit{Cluster 2}} & -1.23                         & \multicolumn{1}{c|}{0.22}           & -1.72                         & \multicolumn{1}{c|}{0.92}           & 2.23                          & \multicolumn{1}{c|}{0.71}           \\
\multicolumn{1}{|l|}{\textit{Cluster 3}} & -2.68                         & \multicolumn{1}{c|}{0.24}           & -1.87                         & \multicolumn{1}{c|}{0.38}           & 3.40                          & \multicolumn{1}{c|}{0.19}           \\
\multicolumn{1}{|l|}{\textit{Cluster 4}} & -0.60                         & \multicolumn{1}{c|}{0.28}           & 0.05                          & \multicolumn{1}{c|}{0.44}           & 0.28                          & \multicolumn{1}{c|}{0.73}           \\
\multicolumn{1}{|l|}{\textit{Cluster 5}} & -2.24                         & \multicolumn{1}{c|}{0.00}           & 2.86                          & \multicolumn{1}{c|}{0.00}           & -1.84                         & \multicolumn{1}{c|}{0.00}           \\ \hline
                                         & \multicolumn{1}{l}{\textbf{}} & \multicolumn{1}{l}{\textbf{}}       & \multicolumn{1}{l}{\textbf{}} & \multicolumn{1}{l}{\textbf{}}       & \multicolumn{1}{l}{\textbf{}} & \multicolumn{1}{l}{\textbf{}}       \\ \cline{2-7} 
                                         & \multicolumn{6}{c}{\textbf{Bing}}                                                                                                                                                                               \\ \cline{2-7} 
\multicolumn{1}{l|}{}                    & \multicolumn{2}{c|}{\textit{\textbf{Black}}}                        & \multicolumn{2}{c|}{\textit{\textbf{Asian}}}                        & \multicolumn{2}{c|}{\textit{\textbf{White}}}                        \\ \cline{2-7} 
\multicolumn{1}{l|}{}                    & \textit{Mean}                 & \multicolumn{1}{c|}{\textit{STDEV}} & \textit{Mean}                 & \multicolumn{1}{c|}{\textit{STDEV}} & \textit{Mean}                 & \multicolumn{1}{c|}{\textit{STDEV}} \\ \hline
\multicolumn{1}{|l|}{Cluster 0}          & -0.53                         & \multicolumn{1}{c|}{0.31}           & -0.66                         & \multicolumn{1}{c|}{0.10}           & 0.90                          & \multicolumn{1}{c|}{0.09}           \\
\multicolumn{1}{|l|}{Cluster 1}          & -1.60                         & \multicolumn{1}{c|}{0.03}           & -0.83                         & \multicolumn{1}{c|}{0.21}           & 1.75                          & \multicolumn{1}{c|}{0.15}           \\
\multicolumn{1}{|l|}{Cluster 2}          & -1.82                         & \multicolumn{1}{c|}{0.02}           & 0.47                          & \multicolumn{1}{c|}{0.74}           & 0.78                          & \multicolumn{1}{c|}{0.44}           \\
\multicolumn{1}{|l|}{Cluster 3}          & -1.33                         & \multicolumn{1}{c|}{0.00}           & 2.70                          & \multicolumn{1}{c|}{0.00}           & -2.37                         & \multicolumn{1}{c|}{0.00}           \\
\multicolumn{1}{|l|}{Cluster 4}          & -2.85                         & \multicolumn{1}{c|}{0.65}           & 0.00                          & \multicolumn{1}{c|}{0.42}           & 2.27                          & \multicolumn{1}{c|}{0.19}           \\
\multicolumn{1}{|l|}{Cluster 5}          & -2.85                         & \multicolumn{1}{c|}{0.22}           & -1.23                         & \multicolumn{1}{c|}{0.05}           & 3.20                          & \multicolumn{1}{c|}{0.23}           \\ \hline
\end{tabular}
\end{scriptsize}
\end{table}

\section{Discussion}
\label{sec:discussion}

In this section, we discuss the stereotypes identified in the previous section.  A positive stereotype exists when the fraction of beautiful women for a given race is larger than the fraction of ugly women for same race. The opposite defines a negative stereotype.

Our results point out  that, for the majority of countries analyzed, there is a positive stereotype for white women and a negative one for black and Asian women. The number of countries for which there is a negative stereotype for black women dominates our statistics, i.e., 85.71\% of the countries collected in Google and 76.47\% in Bing display this type of  stereotype. In the same way we show that there is a negative stereotype about older women. In 69.05\% of the countries in Google and 82.35\% in Bing, the concept of beauty is associated with young women and ugly women are associated with older women. Countries have different configurations of stereotypes, and they can be grouped accordingly. For example, some countries have a very negative stereotype against black women, but can be 'neutral' with respect to the other races.

In the Google dendrogram (Figure~\ref{clusters} (a)), we can highlight cluster 1 - Spain, Guatemala, Argentina, USA, Peru, Mexico, Venezuela, Chile, Brazil and Paraguay - which has a geographical semantic meaning. They are countries from the Americas and Spain. Their population (or a large fraction of it as in the US) speak Latin languages, Spanish and Portuguese. Thus, these are countries with a strong presence of the Hispanic and Latino cultures.  The centroid of this cluster (black: -3.28, Asian: 0.60, white: 2:02) indicates that for this group of countries there is a very negative stereotype regarding black women and a positive stereotype for white women. In Cluster 2 - Ireland, Austria, Germany and Greece - we have European countries and a different stereotype (black: -1.23, Asian: -1.72, white: 2.23) since Asians have a more negative stereotype than blacks. For Cluster 4 - Russia, India, Denmark, Ukraine, South Korea, Kenya, Finland, Japan, Uzbekistan and Afghanistan - we could not identify a clear semantic meaning for the group. However, the cluster has an interesting  stereotype of beauty (black:-0.60, Asian:0.05, white:0.28) in which the stereotype, positive or negative, regarding the races do not exist or are small. There is a coherence between the proportions of the races for the two queries, that is, for most of these countries there is no significant difference between the fractions of the races when we search for beautiful women or ugly women.

In the clustering process of data collected from Bing, Cluster 3 (black:-1.33, Asian:2.70, white:-2.37), composed only by Japan, has the same stereotype of beauty than cluster 5 of Google (black:-2.24, Asian:2.86, white:-1.84), composed only by Malaysia. Both are composed of just an Asian country and therefore have a very positive stereotype regarding Asian and negative stereotype  regarding black women and white women.

In order to deepen the understanding of the stereotypes, we looked at the race composition of some countries to verify if they may explain some of the identified patterns. In Japan, Asians represent 99.4\% of population\footnote{http://www.indexmundi.com/japan/demographics\_profile.html}, in Argentina 97\% of population are white\footnote{http://www.indexmundi.com/argentina/ethnic\_groups.html}, in South Africa 79.2\% are blacks and 8.9\% white\footnote{http://www.southafrica.info/about/people/population.htm\#.V4koMR9yvCI}, at last, in EUA racial composition is 12\% of blacks and 62\% of whites\footnote{http://kff.org/other/state-indicator/distribution-by-raceethnicity/}. Although the racial composition of these countries indicate different fractions of black people, the search engine results show for all of them the presence of the negative stereotype of beauty about black women, with the exception of Japan in Google and Argentina in Bing. We did not find any specific relation between the racial composition of a country and the patterns of stereotypes identified for the country.

\section{Conclusions and Future Work}

To the best of our knowledge, this is the first study to systematically analyze differences in the perception of physical attractiveness of women in the online world. Using a combination of face images obtained by search engine queries plus face’s characteristics inferred by a facial recognition system, the study shows the existence of appearance stereotypes for women in the online world. These findings result from applying a methodology we propose for analyzing stereotypes in online photos that portray people.

Overall, we found negative stereotypes for black and older women. We have demonstrated that this pattern of stereotype is present in almost  all the  continents, Africa,  Asia, Australia/Oceania, Europe, North America, and South America. Our experiments allowed us to pinpoint groups of countries that share similar patterns of stereotypes.  The existence of stereotypes in the online world may foster discrimination both in the online and real world. This is an important contribution of this paper towards actions to reduce bias and discrimination in the online world. 

It is important to emphasize that we do not know exactly the reasons for the existence of the identified  stereotypes. They  may stem from a combination of the stocks of available photos and characteristics of the indexing and ranking algorithms of the search engines.  The stock of photos online may reflect prejudices and bias of the real world that transferred from the physical world to the online world  by the search engines. Given the importance of search engines as  source of information, we suggest that  they analyze the problems caused by the prominent presence of negative stereotypes and find algorithmic ways to minimize the problem.  

Follow-up studies can use mechanical turks to analyze the characteristics of  face images  to generate a more detailed description of classes of stereotypes. Another avenue of future research can  look at the characteristics identified by humans (i.e., mechanical turks) and compare them with the results of different facial recognition systems.

%
%
\bibliographystyle{splncs}
\bibliography{main}

\begin{subappendices}
\renewcommand{\thesection}{\Alph{section}}%

\section{Data gathering statistics}
\label{app1}

Tables \ref{tab1} and \ref{tab2} present the number of photos that Face++ was able to detect a single face per country and for Google and Bing, respectively.

\begin{table}[!h]
\centering
\begin{scriptsize}
\caption{Useful photos from Google.}
\label{tab1}
\begin{tabular}{lcc|lcc|lcc}
\hline
\multicolumn{9}{c}{\textbf{Google}}                                                                                                                                  \\ \hline
\textit{Country} & \textit{Beautiful} & \textit{Ugly} & \textit{Country} & \textit{Beautiful} & \textit{Ugly} & \textit{Country} & \textit{Beautiful} & \textit{Ugly} \\ \hline
Afghanistan      & 37                 & 35            & France           & 37                 & 34            & Morocco          & 30                & 28            \\
South Africa     & 34                 & 38            & Germany          & 29                 & 25            & Nigeria          & 34                & 36            \\
Algeria          & 30                 & 29            & Greece           & 31                 & 27            & Paraguay         & 41                & 33            \\
Angola           & 37                 & 32            & Guatemala        & 41                 & 30            & Peru             & 41                & 30            \\
Saudi Arabia     & 30                 & 30            & India            & 29                 & 35            & Portugal         & 38                & 31            \\
Argentina        & 41                 & 34            & Iraq             & 30                 & 30            & Russia           & 37                & 36            \\
Australia        & 33                 & 38            & Ireland          & 30                 & 22            & Spain            & 40                & 31            \\
Austria          & 39                 & 27            & Italy            & 36                 & 31            & USA              & 35                & 38            \\
Brazil           & 37                 & 30            & Japan            & 39                 & 27            & Sweden           & 46                & 36            \\
Canada           & 33                 & 38            & Kenya            & 34                 & 24            & Turkey           & 40                & 28            \\
Chile            & 39                 & 33            & United Kingdom   & 34                 & 39            & Ukraine          & 41                & 37            \\
Denmark          & 31                 & 24            & South Korea      & 32                 & 33            & Uzbekistan       & 40                & 46            \\
Egypt            & 30                 & 29            & Malaysia         & 39                 & 33            & Venezuela        & 36                & 31            \\
Finland          & 39                 & 25            & Mexico           & 41                 & 32            & Zambia           & 36                & 39           
\end{tabular}
\end{scriptsize}
\end{table}

\begin{table}[!h]
\centering
\begin{scriptsize}
\caption{Useful photos from Bing.}
\label{tab2}
\begin{tabular}{lcc|lcc}
\hline
\multicolumn{6}{c}{\textbf{Bing}}                                                                                             \\ \hline
\textit{Country} & \textit{Beautiful} & \textit{Ugly}        & \textit{Country} & \textit{Beautiful}   & \textit{Ugly}        \\ \hline
South Africa     & 38                 & 44                   & Italy            & 43                   & 40                   \\
Saudi Arabia     & 28                 & \textbf{\textless20} & Japan            & 42                   & 24                   \\
Argentina        & 37                 & 42                   & United Kingdom   & 37                   & 44                   \\
Australia        & 36                 & 45                   & South Korea      & \textbf{\textless20} & \textbf{\textless20} \\
Austria          & 37                 & 34                   & Mexico           & 39                   & 43                   \\
Brazil           & 34                 & 37                   & Portugal         & 32                   & 38                   \\
Canada           & 38                 & 45                   & Russia           & 45                   & 44                   \\
Denmark          & 34                 & 21                   & Spain            & 43                   & 41                   \\
Finland          & 37                 & \textbf{\textless20} & USA              & 37                   & 44                   \\
Germany          & 38                 & 40                   & Turkey           & 42                   & 36                   \\
Greece           & 37                 & \textbf{\textless20} & Ukraine          & 25                   & \textbf{\textless20}
\end{tabular}
\end{scriptsize}
\end{table}

\section{Results of z-score tests}
\label{app2}

In the Figure \ref{zscoreQ1Q2Q3Google} and \ref{zscoreQ1Q2Q3Bing} the results highlighted are those which we reject the null hypothesis and accept the alternative hypothesis. In other words, we can answer YES to the questions {\bf Q1}, {\bf Q2} and/or {\bf Q3}.

\begin{table}[!h]
\centering
\begin{tiny}
\caption{Z-score table associated with the questions Q1, Q2 and Q3 (Google)}
\label{zscoreQ1Q2Q3Google}
\begin{tabular}{lccclccc}
\hline
\multicolumn{8}{c}{\textbf{z-score table (Google)}}                                           \\ \hline
\textit{\textbf{Country}} & \textit{\textbf{Q1 (Black)}}           & \textit{\textbf{Q2 (Asian)}}           & \multicolumn{1}{c}{\textit{\textbf{Q3 (White)}}} & \textit{\textbf{Country}} & \textit{\textbf{Q1 (Black)}}           & \textit{\textbf{Q2 (Asian)}}           & \textit{\textbf{Q3 (White)}}           \\ 

\hline \\[0.0001mm]
Afghanistan               & -0.48                                  & -0.53                                  & \multicolumn{1}{c}{0.74}                         & Italy                     & \cellcolor[HTML]{EFEFEF}\textbf{-2.51} & \cellcolor[HTML]{EFEFEF}\textbf{-1.05} & 2.59                                   \\
South Africa              & \cellcolor[HTML]{EFEFEF}\textbf{-2.96} & -0.34                                  & \multicolumn{1}{c}{2.79}                         & Japan                     & 0.84                                   & 0.62                                   & \cellcolor[HTML]{EFEFEF}\textbf{-0.84} \\
Algeria                   & \cellcolor[HTML]{EFEFEF}\textbf{-2.87} & -0.51                                  & \multicolumn{1}{c}{2.65}                         & Kenya                     & 0.38                                   & -0.13                                  & -0.26                                  \\
Angola                    & \cellcolor[HTML]{EFEFEF}\textbf{-2.99} & -0.19                                  & \multicolumn{1}{c}{2.62}                         & United Kingdom            & \cellcolor[HTML]{EFEFEF}\textbf{-2.91} & 0.14                                   & 2.53                                   \\
Saudi Arabia              & \cellcolor[HTML]{EFEFEF}\textbf{-2.81} & \cellcolor[HTML]{EFEFEF}\textbf{-1.72} & \multicolumn{1}{c}{3.45}                         & South Korea               & \cellcolor[HTML]{EFEFEF}\textbf{-1.41} & 0.65                                   & -0.16                                  \\
Argentina                 & \cellcolor[HTML]{EFEFEF}\textbf{-3.29} & 0.77                                   & \multicolumn{1}{c}{1.82}                         & Malaysia                  & \cellcolor[HTML]{EFEFEF}\textbf{-2.24} & 2.86                                   & \cellcolor[HTML]{EFEFEF}\textbf{-1.84} \\
Australia                 & \cellcolor[HTML]{EFEFEF}\textbf{-2.70} & -0.30                                  & \multicolumn{1}{c}{2.53}                         & Mexico                    & \cellcolor[HTML]{EFEFEF}\textbf{-3.15} & 0.26                                   & 1.98                                   \\
Austria                   & \cellcolor[HTML]{EFEFEF}\textbf{-0.93} & \cellcolor[HTML]{EFEFEF}\textbf{-1.33} & \multicolumn{1}{c}{1.68}                         & Morocco                   & \cellcolor[HTML]{EFEFEF}\textbf{-2.92} & -0.55                                  & 2.73                                   \\
Brazil                    & \cellcolor[HTML]{EFEFEF}\textbf{-3.12} & 0.59                                   & \multicolumn{1}{c}{2.21}                         & Nigeria                   & \cellcolor[HTML]{EFEFEF}\textbf{-2.64} & -0.54                                  & 2.65                                   \\
Canada                    & \cellcolor[HTML]{EFEFEF}\textbf{-2.91} & -0.30                                  & \multicolumn{1}{c}{2.73}                         & Paraguay                  & \cellcolor[HTML]{EFEFEF}\textbf{-3.34} & 1.25                                   & 1.35                                   \\
Chile                     & \cellcolor[HTML]{EFEFEF}\textbf{-3.26} & 0.50                                   & \multicolumn{1}{c}{2.09}                         & Peru                      & \cellcolor[HTML]{EFEFEF}\textbf{-3.26} & 0.16                                   & 2.15                                   \\
Denmark                   & \cellcolor[HTML]{EFEFEF}\textbf{-2.36} & 0.04                                   & \multicolumn{1}{c}{1.50}                         & Portugal                  & \cellcolor[HTML]{EFEFEF}\textbf{-3.10} & -0.26                                  & 2.76                                   \\
Egypt                     & \cellcolor[HTML]{EFEFEF}\textbf{-2.63} & \cellcolor[HTML]{EFEFEF}\textbf{-1.57} & \multicolumn{1}{c}{3.13}                         & Russia                    & \cellcolor[HTML]{EFEFEF}\textbf{-0.89} & -0.03                                  & 0.68                                   \\
Finland                   & 0.21                                   & 0.10                                   & \multicolumn{1}{c}{-0.20}                        & Spain                     & \cellcolor[HTML]{EFEFEF}\textbf{-3.65} & 0.79                                   & 2.01                                   \\
France                    & \cellcolor[HTML]{EFEFEF}\textbf{-2.67} & \cellcolor[HTML]{EFEFEF}\textbf{-1.82} & \multicolumn{1}{c}{3.33}                         & USA                       & \cellcolor[HTML]{EFEFEF}\textbf{-3.20} & 0.66                                   & 2.65                                   \\
Germany                   & \cellcolor[HTML]{EFEFEF}\textbf{-1.59} & \cellcolor[HTML]{EFEFEF}\textbf{-1.06} & \multicolumn{1}{c}{1.97}                         & Sweden                    & \cellcolor[HTML]{EFEFEF}\textbf{-2.88} & \cellcolor[HTML]{EFEFEF}\textbf{-1.09} & 2.79                                   \\
Greece                    & \cellcolor[HTML]{EFEFEF}\textbf{-1.18} & \cellcolor[HTML]{EFEFEF}\textbf{-2.77} & \multicolumn{1}{c}{3.03}                         & Turkey                    & \cellcolor[HTML]{EFEFEF}\textbf{-2.46} & \cellcolor[HTML]{EFEFEF}\textbf{-2.53} & 3.66                                   \\
Guatemala                 & \cellcolor[HTML]{EFEFEF}\textbf{-3.51} & 0.58                                   & \multicolumn{1}{c}{2.15}                         & Ukraine                   & -0.68                                  & 0.58                                   & -0.25                                  \\
India                     & \cellcolor[HTML]{EFEFEF}\textbf{-1.61} & -0.28                                  & \multicolumn{1}{c}{1.07}                         & Uzbekistan                & 0.00                                   & -0.51                                  & 0.51                                   \\
Iraq                      & \cellcolor[HTML]{EFEFEF}\textbf{-2.81} & \cellcolor[HTML]{EFEFEF}\textbf{-1.72} & \multicolumn{1}{c}{3.45}                         & Venezuela                 & \cellcolor[HTML]{EFEFEF}\textbf{-3.01} & 0.43                                   & 1.76                                   \\
Ireland                   & \cellcolor[HTML]{EFEFEF}\textbf{-1.18} & \cellcolor[HTML]{EFEFEF}\textbf{-1.68} & \multicolumn{1}{c}{2.08}                         & Zambia                    & \cellcolor[HTML]{EFEFEF}\textbf{-2.80} & 0.08                                   & 2.43                                  
\end{tabular}
\end{tiny}
\end{table}

\begin{table}[!h]
\centering
\caption{Z-score table associated with the questions Q1, Q2 and Q3 (Bing)}
\label{zscoreQ1Q2Q3Bing}
\begin{tiny}
\begin{tabular}{lccclccc}
\hline
\multicolumn{8}{c}{\textbf{z-score table (Google)}}                                                                                                                                                                                                                                                               \\ \hline
\textit{\textbf{Country}} & \textit{\textbf{Q1 (Black)}}           & \textit{\textbf{Q2 (Asian)}}           & \textit{\textbf{Q3 (White)}} & \textit{\textbf{Country}} & \textit{\textbf{Q1 (Black)}}           & \textit{\textbf{Q2 (Asian)}}           & \textit{\textbf{Q3 (White)}}           \\ \hline
\\[0.0001mm]
South Africa              & \cellcolor[HTML]{EFEFEF}\textbf{-2.86} & \cellcolor[HTML]{EFEFEF}\textbf{-1.28} & 3.23                         & Japan                     & \cellcolor[HTML]{EFEFEF}\textbf{-1.33} & 2.70                                   & \cellcolor[HTML]{EFEFEF}\textbf{-2.37} \\
Argentina                 & -0.69                                  & -0.59                                  & 0.94                         & United Kingdom            & \cellcolor[HTML]{EFEFEF}\textbf{-3.00} & \cellcolor[HTML]{EFEFEF}\textbf{-1.24} & 3.36                                   \\
Australia                 & \cellcolor[HTML]{EFEFEF}\textbf{-2.54} & \cellcolor[HTML]{EFEFEF}\textbf{-1.16} & 2.87                         & Mexico                    & -0.72                                  & -0.59                                  & 0.95                                   \\
Austria                   & -0.06                                  & -0.80                                  & 0.77                         & Portugal                  & \cellcolor[HTML]{EFEFEF}\textbf{-3.32} & 0.29                                   & 2.41                                   \\
Brazil                    & \cellcolor[HTML]{EFEFEF}\textbf{-1.60} & -0.62                                  & 1.62                         & Russia                    & \cellcolor[HTML]{EFEFEF}\textbf{-1.79} & 1.20                                   & 0.31                                   \\
Canada                    & \cellcolor[HTML]{EFEFEF}\textbf{-3.00} & \cellcolor[HTML]{EFEFEF}\textbf{-1.24} & 3.35                         & Spain                     & \cellcolor[HTML]{EFEFEF}\textbf{-1.84} & 0.49                                   & 0.87                                   \\
Denmark                   & \cellcolor[HTML]{EFEFEF}\textbf{-1.83} & -0.27                                  & 1.17                         & USA                       & \cellcolor[HTML]{EFEFEF}\textbf{-2.39} & -0.30                                  & 2.13                                   \\
Germany                   & \cellcolor[HTML]{EFEFEF}\textbf{-1.64} & -0.81                                  & 1.72                         & Turkey                    & \cellcolor[HTML]{EFEFEF}\textbf{-1.57} & \cellcolor[HTML]{EFEFEF}\textbf{-1.05} & 1.91                                   \\
Italy                     & -0.65                                  & -0.65                                  & 0.94                         &                           & \multicolumn{1}{l}{}                   & \multicolumn{1}{l}{}                   & \multicolumn{1}{l}{}                  
\end{tabular}
\end{tiny}
\end{table}

In the Figure \ref{zscoreQ4Q5Q6Google} and \ref{zscoreQ4Q5Q6Bing} the results highlighted are those which we keep the alternative hypothesis and we can answer YES to the questions {\bf Q4}, {\bf Q5} and/or {\bf Q6}.

\begin{table}[!h]
\centering
\caption{Z-score table associated with the questions Q4, Q5 and Q6 (Google)}
\label{zscoreQ4Q5Q6Google}
\begin{tiny}
\begin{tabular}{lccclccc}
\hline
\multicolumn{8}{c}{\textbf{z-score table (Google)}}                                          \\ \hline
\textit{\textbf{Country}} & \textit{\textbf{Q4 (Black)}} & \textit{\textbf{Q5 (Asian)}} & \multicolumn{1}{c|}{\textit{\textbf{Q6 (White)}}}          & \textit{\textbf{Country}} & \textit{\textbf{Q4 (Black)}}          & \textit{\textbf{Q5 (Asian)}}          & \textit{\textbf{Q6 (White)}}          \\ \hline
\\[0.0001mm]
Afghanistan               & -0.48                        & -0.53                        & \multicolumn{1}{c|}{0.74}                                  & Italy                     & -2.51                                 & -1.05                                 & \cellcolor[HTML]{EFEFEF}\textbf{2.59} \\
South Africa              & -2.96                        & -0.34                        & \multicolumn{1}{c|}{\cellcolor[HTML]{EFEFEF}\textbf{2.79}} & Japan                     & \cellcolor[HTML]{EFEFEF}\textbf{0.84} & 0.62                                  & -0.84                                 \\
Algeria                   & -2.87                        & -0.51                        & \multicolumn{1}{c|}{\cellcolor[HTML]{EFEFEF}\textbf{2.65}} & Kenya                     & 0.38                                  & -0.13                                 & -0.26                                 \\
Angola                    & -2.99                        & -0.19                        & \multicolumn{1}{c|}{\cellcolor[HTML]{EFEFEF}\textbf{2.62}} & United Kingdom            & -2.91                                 & 0.14                                  & \cellcolor[HTML]{EFEFEF}\textbf{2.53} \\
Saudi Arabia              & -2.81                        & -1.72                        & \multicolumn{1}{c|}{\cellcolor[HTML]{EFEFEF}\textbf{3.45}} & South Korea               & -1.41                                 & 0.65                                  & -0.16                                 \\
Argentina                 & -3.29                        & 0.77                         & \multicolumn{1}{c|}{\cellcolor[HTML]{EFEFEF}\textbf{1.82}} & Malaysia                  & -2.24                                 & \cellcolor[HTML]{EFEFEF}\textbf{2.86} & -1.84                                 \\
Australia                 & -2.70                        & -0.30                        & \multicolumn{1}{c|}{\cellcolor[HTML]{EFEFEF}\textbf{2.53}} & Mexico                    & -3.15                                 & 0.26                                  & \cellcolor[HTML]{EFEFEF}\textbf{1.98} \\
Austria                   & -0.93                        & -1.33                        & \multicolumn{1}{c|}{\cellcolor[HTML]{EFEFEF}\textbf{1.68}} & Morocco                   & -2.92                                 & -0.55                                 & \cellcolor[HTML]{EFEFEF}\textbf{2.73} \\
Brazil                    & -3.12                        & 0.59                         & \multicolumn{1}{c|}{\cellcolor[HTML]{EFEFEF}\textbf{2.21}} & Nigeria                   & -2.64                                 & -0.54                                 & \cellcolor[HTML]{EFEFEF}\textbf{2.65} \\
Canada                    & -2.91                        & -0.30                        & \multicolumn{1}{c|}{\cellcolor[HTML]{EFEFEF}\textbf{2.73}} & Paraguay                  & -3.34                                 & \cellcolor[HTML]{EFEFEF}\textbf{1.25} & \cellcolor[HTML]{EFEFEF}\textbf{1.35} \\
Chile                     & -3.26                        & 0.50                         & \multicolumn{1}{c|}{\cellcolor[HTML]{EFEFEF}\textbf{2.09}} & Peru                      & -3.26                                 & 0.16                                  & \cellcolor[HTML]{EFEFEF}\textbf{2.15} \\
Denmark                   & -2.36                        & 0.04                         & \multicolumn{1}{c|}{\cellcolor[HTML]{EFEFEF}\textbf{1.50}} & Portugal                  & -3.10                                 & -0.26                                 & \cellcolor[HTML]{EFEFEF}\textbf{2.76} \\
Egypt                     & -2.63                        & -1.57                        & \multicolumn{1}{c|}{\cellcolor[HTML]{EFEFEF}\textbf{3.13}} & Russia                    & -0.89                                 & -0.03                                 & 0.68                                  \\
Finland                   & 0.21                         & 0.10                         & \multicolumn{1}{c|}{-0.20}                                 & Spain                     & -3.65                                 & 0.79                                  & \cellcolor[HTML]{EFEFEF}\textbf{2.01} \\
France                    & -2.67                        & -1.82                        & \multicolumn{1}{c|}{\cellcolor[HTML]{EFEFEF}\textbf{3.33}} & USA                       & -3.20                                 & 0.66                                  & \cellcolor[HTML]{EFEFEF}\textbf{2.65} \\
Germany                   & -1.59                        & -1.06                        & \multicolumn{1}{c|}{\cellcolor[HTML]{EFEFEF}\textbf{1.97}} & Sweden                    & -2.88                                 & -1.09                                 & \cellcolor[HTML]{EFEFEF}\textbf{2.79} \\
Greece                    & -1.18                        & -2.77                        & \multicolumn{1}{c|}{\cellcolor[HTML]{EFEFEF}\textbf{3.03}} & Turkey                    & -2.46                                 & -2.53                                 & \cellcolor[HTML]{EFEFEF}\textbf{3.66} \\
Guatemala                 & -3.51                        & 0.58                         & \multicolumn{1}{c|}{\cellcolor[HTML]{EFEFEF}\textbf{2.15}} & Ukraine                   & -0.68                                 & 0.58                                  & -0.25                                 \\
India                     & -1.61                        & -0.28                        & \multicolumn{1}{c|}{\cellcolor[HTML]{EFEFEF}\textbf{1.07}} & Uzbekistan                & 0.00                                  & -0.51                                 & 0.51                                  \\
Iraq                      & -2.81                        & -1.72                        & \multicolumn{1}{c|}{\cellcolor[HTML]{EFEFEF}\textbf{3.45}} & Venezuela                 & -3.01                                 & 0.43                                  & \cellcolor[HTML]{EFEFEF}\textbf{1.76} \\
Ireland                   & -1.18                        & -1.68                        & \multicolumn{1}{c|}{\cellcolor[HTML]{EFEFEF}\textbf{2.08}} & Zambia                    & -2.80                                 & 0.08                                  & \cellcolor[HTML]{EFEFEF}\textbf{2.43}
\end{tabular}
\end{tiny}
\end{table}

\begin{table}[!h]
\centering
\caption{Z-score table associated with the questions Q4, Q5 and Q6 (Bing)}
\label{zscoreQ4Q5Q6Bing}
\begin{tiny}
\begin{tabular}{lccclccc}
\hline
\multicolumn{8}{c}{\textbf{z-score table (Bing)}}                                                                                                                                                                                                                                         \\ \hline
\textit{\textbf{Country}} & \textit{\textbf{Q1 (Black)}} & \textit{\textbf{Q2 (Asian)}} & \textit{\textbf{Q3 (White)}}          & \textit{\textbf{Country}} & \textit{\textbf{Q1 (Black)}} & \textit{\textbf{Q2 (Asian)}}          & \textit{\textbf{Q3 (White)}}          \\ \hline
\\[0.0001mm]
South Africa              & -2.86                 .       & -1.28                        & \cellcolor[HTML]{EFEFEF}\textbf{3.23} & Japan                     & -1.33                        & \cellcolor[HTML]{EFEFEF}\textbf{2.70} & -2.37                                 \\
Argentina                 & -0.69                        & -0.59                        & \cellcolor[HTML]{EFEFEF}\textbf{0.94} & United Kingdom            & -3.00                        & -1.24                                 & \cellcolor[HTML]{EFEFEF}\textbf{3.36} \\
Australia                 & -2.54                        & -1.16                        & \cellcolor[HTML]{EFEFEF}\textbf{2.87} & Mexico                    & -0.72                        & -0.59                                 & \cellcolor[HTML]{EFEFEF}\textbf{0.95} \\
Austria                   & -0.06                        & -0.80                        & 0.77                                  & Portugal                  & -3.32                        & 0.29                                  & \cellcolor[HTML]{EFEFEF}\textbf{2.41} \\
Brazil                    & -1.60                        & -0.62                        & \cellcolor[HTML]{EFEFEF}\textbf{1.62} & Russia                    & -1.79                        & \cellcolor[HTML]{EFEFEF}\textbf{1.20} & 0.31                                  \\
Canada                    & -3.00                        & -1.24                        & \cellcolor[HTML]{EFEFEF}\textbf{3.35} & Spain                     & -1.84                        & 0.49                                  & \cellcolor[HTML]{EFEFEF}\textbf{0.87} \\
Denmark                   & -1.83                        & -0.27                        & \cellcolor[HTML]{EFEFEF}\textbf{1.17} & USA                       & -2.39                        & -0.30                                 & \cellcolor[HTML]{EFEFEF}\textbf{2.13} \\
Germany                   & -1.64                        & -0.81                        & \cellcolor[HTML]{EFEFEF}\textbf{1.72} & Turkey                    & -1.57                        & -1.05                                 & \cellcolor[HTML]{EFEFEF}\textbf{1.91} \\
Italy                     & -0.65                        & -0.65                        & \cellcolor[HTML]{EFEFEF}\textbf{0.94} &                           & \multicolumn{1}{l}{}         & \multicolumn{1}{l}{}                  & \multicolumn{1}{l}{}                 
\end{tabular}
\end{tiny}
\end{table}

\section{Results of Wilcoxon tests}
\label{app3}

Results highlighted in the Tables~\ref{pvalueQ7Google} and \ref{pvalueQ7Bing} show those countries for which we keep the alternative hypothesis.

\begin{table}[!h]
\centering
\begin{scriptsize}
\caption{P-value table associated with the questions {\bf Q7} (Google)}
\label{pvalueQ7Google}
\begin{tabular}{lclclc}
\hline
\multicolumn{6}{c}{\textbf{Google}}                                                                                                                                                                                            \\ \hline
\multicolumn{6}{c}{\textit{\textbf{Wilcoxon test (Q7)}}}                                                                                                                                                                       \\ \hline
\textit{Country} & \multicolumn{1}{c|}{\textit{p-value}}                        & \textit{Country} & \multicolumn{1}{c|}{\textit{p-value}}                        & \textit{Country} & \textit{p-value}                        \\ \hline
Afghanistan      & \multicolumn{1}{c|}{0.1819}                                  & France           & \multicolumn{1}{c|}{0.0572}                                  & Morocco          & \cellcolor[HTML]{EFEFEF}\textbf{0.0036} \\
South Africa     & \multicolumn{1}{c|}{\cellcolor[HTML]{EFEFEF}\textbf{0.0001}} & Germany          & \multicolumn{1}{c|}{\cellcolor[HTML]{EFEFEF}\textbf{0.0107}} & Nigeria          & \cellcolor[HTML]{EFEFEF}\textbf{0.0000} \\
Algeria          & \multicolumn{1}{c|}{\cellcolor[HTML]{EFEFEF}\textbf{0.0023}} & Greece           & \multicolumn{1}{c|}{\cellcolor[HTML]{EFEFEF}\textbf{0.0040}} & Paraguay         & \cellcolor[HTML]{EFEFEF}\textbf{0.0471} \\
Angola           & \multicolumn{1}{c|}{\cellcolor[HTML]{EFEFEF}\textbf{0.0072}} & Guatemala        & \multicolumn{1}{c|}{0.0512}                                  & Peru             & \cellcolor[HTML]{EFEFEF}\textbf{0.0499} \\
Saudi Arabia     & \multicolumn{1}{c|}{\cellcolor[HTML]{EFEFEF}\textbf{0.0131}} & India            & \multicolumn{1}{c|}{0.1221}                                  & Portugal         & \cellcolor[HTML]{EFEFEF}\textbf{0.0014} \\
Argentina        & \multicolumn{1}{c|}{\cellcolor[HTML]{EFEFEF}\textbf{0.0271}} & Iraq             & \multicolumn{1}{c|}{\cellcolor[HTML]{EFEFEF}\textbf{0.0196}} & Russia           & \cellcolor[HTML]{EFEFEF}\textbf{0.0146} \\
Australia        & \multicolumn{1}{c|}{\cellcolor[HTML]{EFEFEF}\textbf{0.0003}} & Ireland          & \multicolumn{1}{c|}{0.0703}                                  & Spain            & \cellcolor[HTML]{EFEFEF}\textbf{0.1869} \\
Austria          & \multicolumn{1}{c|}{\cellcolor[HTML]{EFEFEF}\textbf{0.0017}} & Italy            & \multicolumn{1}{c|}{0.2288}                                  & USA              & \cellcolor[HTML]{EFEFEF}\textbf{0.0000} \\
Brazil           & \multicolumn{1}{c|}{\cellcolor[HTML]{EFEFEF}\textbf{0.0298}} & Japan            & \multicolumn{1}{c|}{0.0520}                                  & Sweden           & \cellcolor[HTML]{EFEFEF}\textbf{0.0071} \\
Canada           & \multicolumn{1}{c|}{\cellcolor[HTML]{EFEFEF}\textbf{0.0001}} & Kenya            & \multicolumn{1}{c|}{\cellcolor[HTML]{EFEFEF}\textbf{0.0041}} & Turkey           & \cellcolor[HTML]{EFEFEF}\textbf{0.0093} \\
Chile            & \multicolumn{1}{c|}{\cellcolor[HTML]{EFEFEF}\textbf{0.0134}} & United Kingdom   & \multicolumn{1}{c|}{\cellcolor[HTML]{EFEFEF}\textbf{0.0000}} & Ukraine          & 0.1699                                  \\
Denmark          & \multicolumn{1}{c|}{0.3731}                                  & South Korea      & \multicolumn{1}{c|}{0.1363}                                  & Uzbekistan       & 0.8407                                  \\
Egypt            & \multicolumn{1}{c|}{\cellcolor[HTML]{EFEFEF}\textbf{0.0122}} & Malaysia         & \multicolumn{1}{c|}{\cellcolor[HTML]{EFEFEF}\textbf{0.0005}} & Venezuela        & \cellcolor[HTML]{EFEFEF}\textbf{0.0218} \\
Finland          & \multicolumn{1}{c|}{0.1759} & Mexico         & \multicolumn{1}{c|}{\cellcolor[HTML]{EFEFEF}\textbf{0.0174}} & Zambia        & \cellcolor[HTML]{EFEFEF}\textbf{0.0002} 
\end{tabular}
\end{scriptsize}
\end{table}

\begin{table}[!h]
\centering
\caption{P-value table associated with the questions {\bf Q7} (Bing)}
\label{pvalueQ7Bing}
\begin{scriptsize}
\begin{tabular}{lclc}
\hline
\multicolumn{4}{c}{\textbf{Bing}}                                                                                                            \\ \hline
\multicolumn{4}{c}{\textit{\textbf{Wilcoxon test (Q7)}}}                                                                                     \\ \hline
\textit{Country} & \multicolumn{1}{c|}{\textit{p-value}}                        & \textit{Country} & \textit{p-value}                        \\ \hline
South Africa     & \multicolumn{1}{c|}{\cellcolor[HTML]{EFEFEF}\textbf{0.0179}} & Japan            & 0.1058                                  \\
Argentina        & \multicolumn{1}{c|}{0.0612}                                  & United Kingdom   & \cellcolor[HTML]{EFEFEF}\textbf{0.0226} \\
Australia        & \multicolumn{1}{c|}{\cellcolor[HTML]{EFEFEF}\textbf{0.0077}} & Mexico           & \cellcolor[HTML]{EFEFEF}\textbf{0.0257} \\
Austria          & \multicolumn{1}{c|}{\cellcolor[HTML]{EFEFEF}\textbf{0.0001}} & Portugal         & \cellcolor[HTML]{EFEFEF}\textbf{0.0314} \\
Brazil           & \multicolumn{1}{c|}{\cellcolor[HTML]{EFEFEF}\textbf{0.0002}} & Russia           & \cellcolor[HTML]{EFEFEF}\textbf{0.0302} \\
Canada           & \multicolumn{1}{c|}{\cellcolor[HTML]{EFEFEF}\textbf{0.0211}} & Spain            & 0.0553                                  \\
Denmark          & \multicolumn{1}{c|}{\cellcolor[HTML]{EFEFEF}\textbf{0.0168}} & USA              & \cellcolor[HTML]{EFEFEF}\textbf{0.0021} \\
Germany          & \multicolumn{1}{c|}{\cellcolor[HTML]{EFEFEF}\textbf{0.0012}} & Turkey           & \cellcolor[HTML]{EFEFEF}\textbf{0.0040} \\
Italy            & \multicolumn{1}{c|}{\cellcolor[HTML]{EFEFEF}\textbf{0.0025}} &                  &                                        
\end{tabular}
\end{scriptsize}
\end{table}
\end{subappendices}

\end{document}